\documentclass[%
superscriptaddress,
twocolumn,
showpacs,
amsmath,amssymb,
aps,
prl,
floatfix,
showkeys,
reprint,
]{revtex4-1}

\usepackage{graphicx}
\usepackage{dcolumn}
\usepackage{bm}
\usepackage{amsmath}
\usepackage{amssymb}
\usepackage{multirow}
\usepackage{color}
\usepackage{booktabs}
\usepackage{tabularx}
\usepackage[english]{babel}
\usepackage[utf8]{inputenc}
\usepackage{enumerate}
\usepackage[cmyk,dvipsnames]{xcolor}
\definecolor{pacificb}{HTML}{1CA9C9}

\usepackage[normalem]{ulem}

\begin{document}

\title{Discovery of magnetic single- and triple-Q states in Mn/Re(0001)}

\author{Jonas Spethmann}
\affiliation{Department of Physics, University of Hamburg, 20355 Hamburg, Germany}
\author{Sebastian Meyer}
\affiliation{Institute of Theoretical Physics and Astrophysics, University of Kiel, Leibnizstrasse 15, 24098 Kiel, Germany}
\author{Kirsten von Bergmann}
\affiliation{Department of Physics, University of Hamburg, 20355 Hamburg, Germany}
\author{Roland Wiesendanger}
\affiliation{Department of Physics, University of Hamburg, 20355 Hamburg, Germany}
\author{Stefan Heinze}
\affiliation{Institute of Theoretical Physics and Astrophysics, University of Kiel, Leibnizstrasse 15, 24098 Kiel, Germany}
\author{Andr\'e Kubetzka}
\email[Corresp. author: ]{kubetzka@physnet.uni-hamburg.de}
\affiliation{Department of Physics, University of Hamburg, 20355 Hamburg, Germany}
	   
\date{\today}

\begin{abstract}
We experimentally verify the existence of two model-type magnetic ground states which were previously predicted but so far unobserved. We find them in Mn monolayers on the Re(0001) surface using spin-polarized scanning tunneling microscopy. For fcc stacking of Mn the collinear row-wise antiferromagnetic state occurs, whereas for hcp Mn a three-dimensional spin structure appears, which is a superposition of three row-wise antiferromagnetic states and is known as triple-Q state. Density functional theory calculations elucidate the subtle interplay of different magnetic interactions to form these spin structures and provide insight into the role played by relativistic effects.
\noindent
\end{abstract}


\maketitle

In nano-scale non-collinear magnetic systems various magnetic interactions can compete ranging from isotropic Heisenberg exchange over spin-orbit coupling (SOC) related effects such as the Dzyaloshinskii-Moriya interaction to recently proposed interactions arising from topological orbital moments. The Heisenberg exchange interaction is described by a symmetric bilinear term of the form $-J_{ij}(\mathbf{S}_i \cdot \mathbf{S}_j)$, with the exchange constant $J_{ij}$ giving the strength of the interaction between spins $\mathbf{S}_i$ and $\mathbf{S}_j$. From the general form of the exchange tensor other pairwise interactions can be derived, namely the antisymmetric Dzyaloshinskii-Moriya interaction (DMI) and the anisotropic symmetric exchange (ASE) which both are relativistic effects arising from spin-orbit coupling~\cite{dzyaloshinskiiSPJETP1957,moriyaPR1960,smithJMMM1976,stauntonJPCSSP1988}. The DMI has the form $-\mathbf{D}_{ij} (\mathbf{S}_i \times \mathbf{S}_j)$ and thus favors non-collinear magnetic order with unique rotational sense. The ASE can be written as $-J^\mathrm{ASE}_{ij}(\mathbf{S}_i \cdot \mathbf{d}_{ij})(\mathbf{S}_j \cdot \mathbf{d}_{ij})$, where $\mathbf{d}_{ij}$ is the unit vector pointing from $\mathbf{S}_i$ to $\mathbf{S}_j$~\cite{smithJMMM1976}; because this term describes the anisotropic part of dipolar interactions the ASE is also referred to as pseudo-dipolar interaction. Whereas DMI induced non-collinear magnetic order has been in the focus of recent research on magnetic domain walls and skyrmions~\cite{vonbergmannJPCM2014,wiesendangerNRM2016,fertNRM2017}, the ASE is rarely taken into account to model experimental systems~\cite{hermenauNC2019}. 

Beyond the pairwise magnetic exchange couplings also higher-order interactions (HOI) between four spins have been considered \cite{hoffmannPRB2020}, and in several systems their importance for the magnetic ground state has been demonstrated in combined experimental and theoretical studies~\cite{heinzeNP2011,yoshidaPRL2012,rommingPRL2018,kronleinPRL2018}. Recently, additional interactions have been proposed for transition metals, e.g.\ higher-order DMI~\cite{mankovskyAC2019,brinkerNJP2019} and interactions involving topological orbital moments arising when the solid angle of three adjacent spins is non-zero~\cite{hoffmannPRB2015,diasNC2016,hankePRB2016,diasNC2016,grytsiukNC2020}.

When magnetic interactions compete, even structurally simple systems can host complex magnetic states with exciting new properties. In this respect hexagonal magnetic monolayers can serve as auspicious model-type systems. For such a symmetry, antiferromagnetic (AFM) nearest neighbor exchange coupling leads to geometric frustration and the ground state is 
a N\'eel state with $120^{\circ}$ between adjacent magnetic moments~\cite{wortmannPRL2001,gaoPRL2008b,ouaziPRL2014}. When exchange interactions beyond nearest neighbor AFM exchange play a role, e.g.\ when $1 > J_2/J_1 > 1/8$, the row-wise antiferromagnetic (RW-AFM) state can arise~\cite{kurzPRL2001,hardratPRB2009}. The RW-AFM and the N\'eel state can both be expressed as spin spirals characterized by a single spin spiral vector $\mathbf q$.

Within the Heisenberg model the RW-AFM state is degenerate with the so-called triple-$\mathbf q$ (3Q) state, which can be constructed by a superposition of three symmetry-equivalent RW-AFM states.
However, HOIs can lift this degeneracy and favor one state over the other. Nearly two decades ago the 3Q state was predicted based on DFT calculations for a Mn monolayer on Cu(111)~\cite{kurzPRL2001}, but the experimental realization of this system suffered from severe intermixing. Recent calculations for an unsupported Mn layer with a 3Q state indicate sizable topological orbital moments~\cite{hankePRB2016} which could lead to additional chiral-chiral and spin-chiral interactions~\cite{grytsiukNC2020}. To the best of our knowledge neither the RW-AFM state nor the 3Q state have been discovered experimentally up to now.

Here we study the magnetic ground state of Mn monolayers on Re(0001). Using spin-polarized (SP-) STM we find that the RW-AFM state occurs in fcc-stacked Mn whereas the hcp-stacked Mn exhibits the 3Q state. DFT calculations show large values for the HOIs, however, because of the competition between different HOIs the single-$\mathbf q$ RW-AFM state and the 3Q state are nearly degenerate. The experiments show preferred orientations of both magnetic states with respect to the crystallographic directions. To understand the origin of this coupling we consider different energy contributions such as dipole-dipole interaction and spin-orbit induced ASE.

\begin{figure}[tb]
	\centering
	\includegraphics[width=0.9\columnwidth]{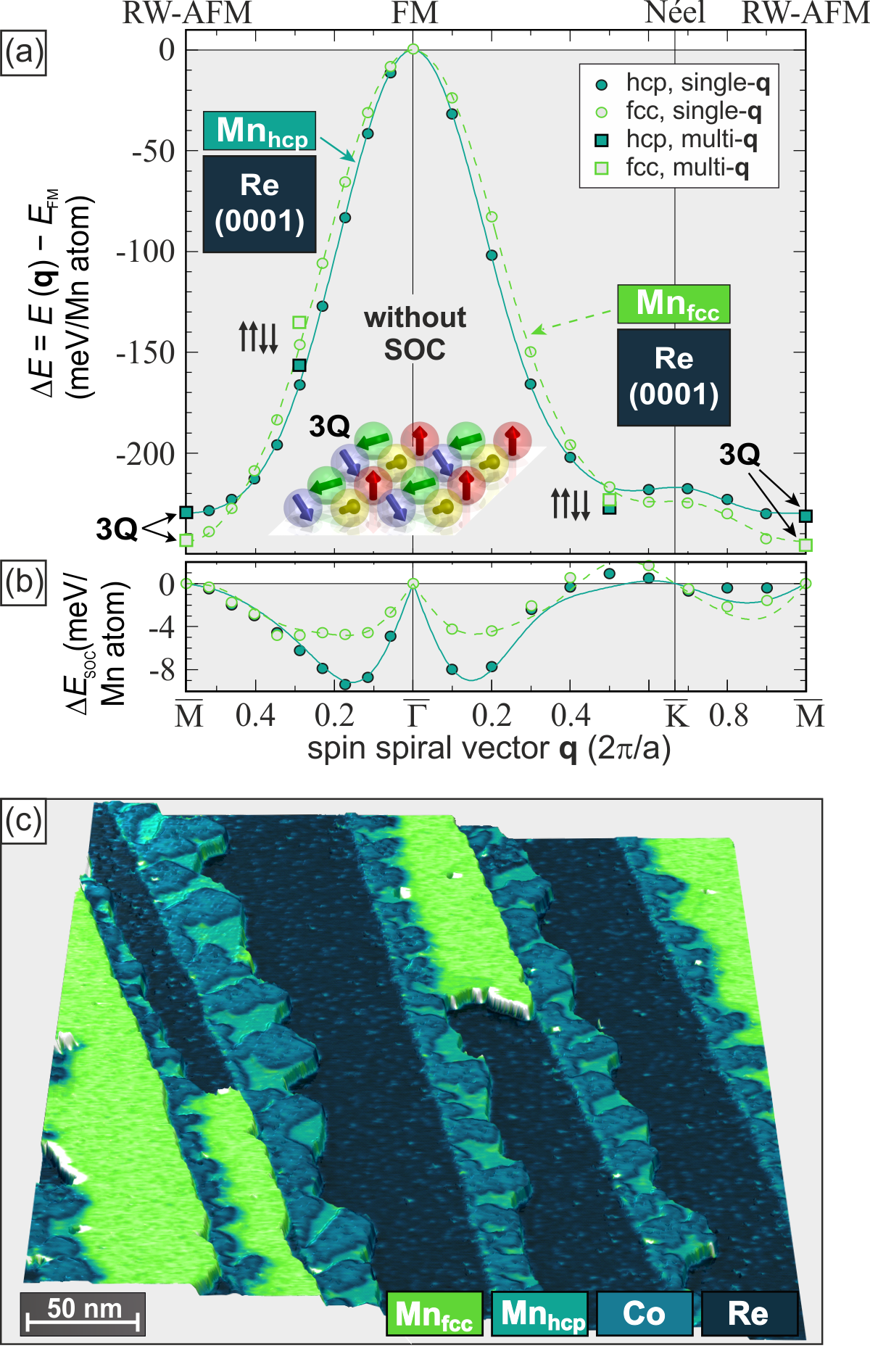}
	\caption{\textbf{(a)}~Energy dispersion $E(\mathbf{q})$ of spin spirals obtained via DFT along the two high symmetry directions of the two-dimensional Brillouin zone
  for both stackings of Mn on Re(0001) without spin-orbit coupling. The energies of three multi-$\mathbf{q}$ states (3Q and two $\uparrow\uparrow\downarrow\downarrow$) are indicated at the
  $\mathbf{q}$ vector of the corresponding single-$\mathbf{q}$ state.
  \textbf{(b)}~Calculated energy contribution to spin spirals due to spin-orbit coupling. \textbf{(c)}~Perspective view of a constant-current STM topography image of Mn on Re(0001) colorized with the simultaneously acquired differential conductance (d$I$/d$U$)~signal. Co, decorating the Re step edges, has induced hcp Mn growth; $U=+500$\,mV, $I=1.2$\,nA, $T=4$\,K.}
	\label{fig1}
\end{figure}

In order to scan a large part of the magnetic phase space, we calculate via DFT the energy dispersion $E(\mathbf{q})$ of spin spiral states, Fig.\,\ref{fig1}(a). For a spin spiral characterized by $\mathbf{q}$ the magnetic moment of an atom at site $\mathbf{R}_i$ is given by $\mathbf{M}_i=M (\cos{\mathbf{q} \cdot \mathbf{R}_i}, \sin{\mathbf{q} \cdot \mathbf{R}_i}, 0)$ where $M$ is the magnetic moment. For Mn/Re(0001) we find $M_{\mathrm Mn} \approx 3.3 \mu_{\rm B}$, a value nearly independent of $\mathbf{q}$. The ferromagnetic (FM) state at the $\overline{\Gamma}$-point has a much higher energy compared to the antiferromagnetic $120^\circ$ N\'eel state ($\overline{\mathrm K}$-point) and the RW-AFM state ($\overline{\mathrm M}$-point).
For both the fcc- and the hcp-stacked Mn monolayer the RW-AFM state is the lowest energy state of all single-$\mathbf q$ states and the fcc stacking of Mn is preferred over hcp Mn by 27.4\,meV/Mn atom~\cite{suppMnRe}. By mapping these DFT energy dispersions to the Heisenberg model we obtain the exchange constants. We find that both nearest-neighbor and next-nearest neighbor coupling, $J_1$ and $J_2$, are antiferromagnetic, with a ratio expected for a RW-AFM state~\cite{hardratPRB2009}, see Table\,I and Ref.~\cite{suppMnRe}.

To elucidate whether a 3Q state can occur in Mn/Re(0001) we calculate its total energy with respect to the RW-AFM state. We find that the 3Q state is slightly favored for both stackings (see Table\,I), however, the small energy differences do not indicate that the HOIs are negligible. We can determine the strength of the two-site ($B_1$), three-site ($Y_1$), and four-site ($K_1$) four spin interaction by calculating in addition the total energy of the two different double-row wise AFM ($\uparrow\uparrow\downarrow\downarrow$) states~\cite{hardratPRB2009,rommingPRL2018,kronleinPRL2018,hoffmannPRB2020} with respect to the corresponding single-$\mathbf{q}$ states~\cite{suppMnRe}, i.e.~$90^\circ$ spin spirals, see Fig.\,\ref{fig1}(a). The obtained values of $B_1$, $Y_1$, and $K_1$ are of significant strength, see Table\,I, but their net contribution to the energy of 3Q and RW-AFM state nearly cancels.

\begin{table}
\centering
\caption {Calculated values (in meV) for Heisenberg exchange constants $J^{\prime}_1$ to $J^{\prime}_3$, where the prime
denotes that the effect of higher-order exchange interactions is taken into account in the fit of the energy dispersion,
the higher-order exchange constants $B_1$, $Y_1$, and $K_1$ and the energy difference 
$\Delta E=E_\text{3Q}-E_\text{RW-AFM}$ in meV/Mn atom~\cite{suppMnRe}.}
\begin{ruledtabular}
\begin{tabular}{lccccccc} 
 & $J^{\prime}_1$ & $J^{\prime}_2$ & $J^{\prime}_3$ & $B_1$ & $Y_1$ & $K_1$ & $\Delta E$ \\
 \hline 
fcc & $-22.4$ & $-3.4$ & $0.88$ & $-1.56$ & $-2.29$ & $-0.43$ & $-0.7$ \\
hcp & $-18.7$ & $-4.2$ & $-1.38$ & $-1.25$ & $-2.49$ & $-0.66$ & $-0.4$ 
\end{tabular}
\end{ruledtabular}
\label{table}
\end{table}

Spin-orbit coupling effects might also contribute to the formation of the magnetic ground state. We find for both stackings of the Mn an easy-plane magnetocrystalline anisotropy energy (MAE) on the order of 1\,meV/Mn atom. The calculated energy contribution due to SOC for cylcoidal spin spirals is shown in Fig.\,\ref{fig1}(b). We find that it is large near $\overline{\Gamma}$, which corresponds to the large DMI reported previously~\cite{belabbesPRL2016}. However, its impact is significantly reduced at $\overline{\mathrm M}$, i.e.~close to the RW-AFM state.
Consequently, it is necessary to go beyond nearest neighbor contributions to capture the dispersion of the DMI energy~\cite{suppMnRe}. So far, the DFT results demonstrate a competition of different types of interactions preventing a robust prediction of the magnetic ground states.

In the experiment, Mn grows on Re(0001) almost exclusively in fcc-stacking~\cite{suppMnRe}, in agreement with the DFT calculations. Extended areas of hcp Mn can be induced by previously growing hcp Co which decorates the Re step edges, as can be seen in Fig.~\ref{fig1}(c), where hcp and fcc Mn areas coexist. 
\begin{figure}[tb]
	\centering
	\includegraphics[width=1\columnwidth]{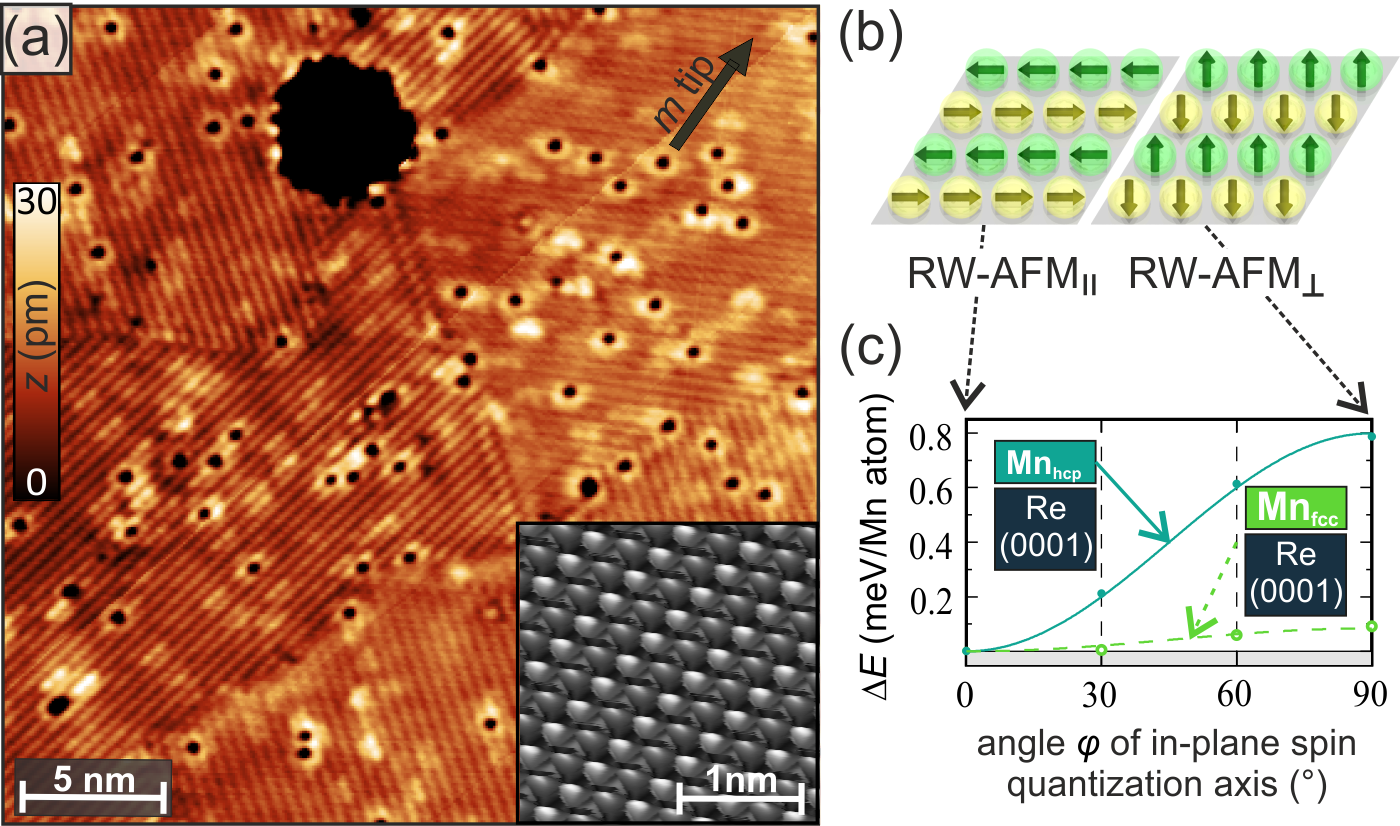}%
	\caption{\textbf{(a)}~Spin-resolved constant-current STM image of the fcc-stacked Mn monolayer showing three types of rotational domains of the RW-AFM state; Fe-coated W tip, $U=-20$\,mV, $I=7.5$\,nA, $T=8$\,K, $B=0$\,T, raw data. The atom manipulation image (inset) demonstrates commensurablity of magnetic state and atomic lattice (Co adatom, Cr tip, $U=5$\,mV, $I=4$\,nA). \textbf{(b)}~Spin structures of RW-AFM states with spin quantization axes parallel and perpendicular to the rows. \textbf{(c)}~DFT calculation of the energy of in-plane RW-AFM states as a function of the angle of the spin quantization axis.}
	\label{fig2}
\end{figure}
Figure~\ref{fig2}(a) shows a closer view of an fcc Mn monolayer area, grown without Co. Here, a spin-sensitive Fe-coated W tip is used, which is typically sensitive to the in-plane components of the sample magnetization in zero field~\cite{bodeRPP2003}. We observe three rotational domains of stripes running along the close-packed rows. 
The distance between the stripes is exactly two atomic rows as inferred from magnetic atom manipulation imaging (see inset and Ref.~\cite{suppMnRe}) and thus we conclude that fcc Mn exhibits the RW-AFM state. Different rotational domains can show different magnetic contrast amplitudes, e.g.\ the contrast is lowest for the domain in the upper right. An out-of-plane easy axis would result in the same contrast independent of the rotational domain and can therefore be excluded, in agreement with the easy-plane MAE obtained from DFT. Furthermore, we find a strong correlation of domain angle and magnetic corrugation amplitude -- in this data and in general -- with almost no exceptions.
This means, that the spin direction is coupled to the magnetic rows of the AFM state in one of the two ways depicted in Fig.\,\ref{fig2}(b).

Surprisingly, none of the previously considered interactions, i.e.\ Heisenberg exchange, HOI, MAE, or DMI, can mediate this kind of coupling.
However, so far we have -- as often reasonable for ultrathin films and antiferromagnets -- neglected the magnetic dipole-dipole interaction. We can calculate its energy contribution for the two  states based on the magnetic moment of 3.3\,$\mu_B$ per Mn atom as obtained from DFT. Thereby, we find that the configuration with spin quantization axis parallel to the rows, RW-AFM$_\|$, is favored by 0.14~meV/Mn atom compared to RW-AFM$_\perp$, a value roughly $2/3$ of the shape anisotropy of a FM state.

In addition to the dipolar contribution, the spin-orbit coupling induced ASE can lead to an energy difference between the two configurations. We can quantify this
effect for both stackings by calculating the energy of the RW-AFM state for different rotations of the spin quantization axis with respect to the direction of the rows, see Fig.\,\ref{fig2}(c). We find that in fcc Mn the ASE leads to an energy difference of about 0.1 meV/Mn atom, with a preference for the RW-AFM$_\|$ state.
Both effects, the dipole-dipole interaction and the ASE are thus of similar strength and mediate the same orientation of spins in fcc Mn/Re(0001).

\begin{figure}[tb]
	\centering
	\includegraphics[width=1\columnwidth]{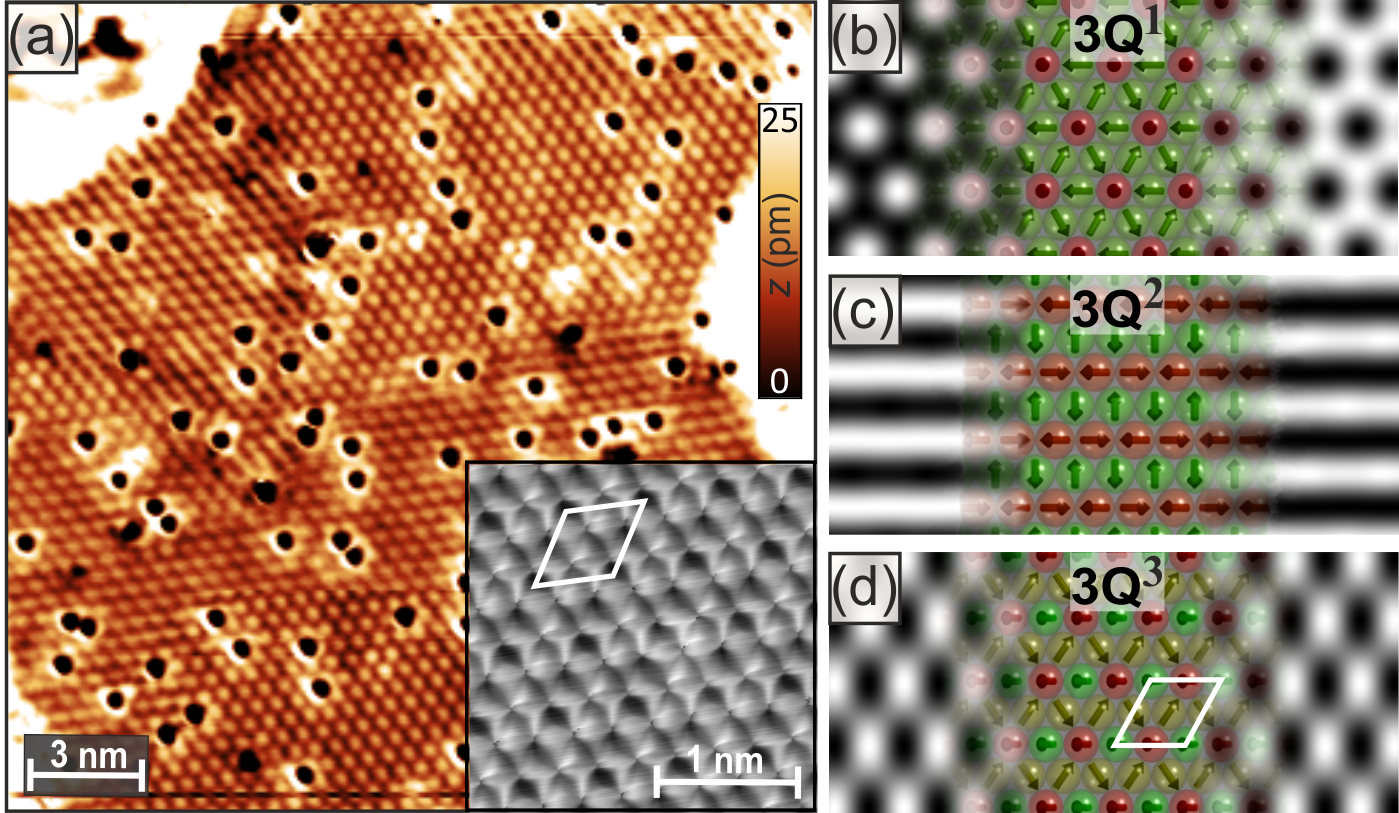}%
	\caption{\textbf{(a)}~Spin-resolved constant-current STM image of the hcp-stacked Mn monolayer showing three types of rotational domains of the 3Q$^3$ state; Cr tip, $U=-30$\,mV, $I=7$\,nA, $T=4$\,K, $B=0$\,T, raw data. In the inset the commensurability of the 3Q state is demonstrated by atom manipulation (Co adatom, Cr tip, $U=5$\,mV, $I=10$\,nA). \textbf{(b)},\textbf{(c)},\textbf{(d)}~Spin structures (red: up, green: down)	and SP-STM simulations of three differently oriented 3Q$^i$ states, with tip magnetization pointing up (left side) and down (right side).}
	\label{fig3}
\end{figure}

Figure~3(a) shows a spin-resolved STM image of hcp Mn monolayer and adjacent hcp Co areas (white). We observe a hexagonal superstructure with twice the atomic lattice constant, i.e.\ 4 atoms in the magnetic unit cell as found from magnetic atom manipulation imaging (see inset and Ref.~\cite{suppMnRe}), compatible with the presence of a 3Q state. In different regions of the Mn monolayer the details of the hexagonal pattern change and we find three qualitatively different regions in Fig.\,3(a), indicating rotational domains analogous to the RW-AFM domains in Fig.\,2(a).
Since SP-STM is sensitive to the projection of surface spins onto the tip magnetization~\cite{bodeRPP2003}, the observed patterns depend on the 3Q rotational domain as well as the tip magnetization direction~\cite{suppMnRe,wortmannJMMM2002,heinzeAPA2006}. 

\begin{figure*}[tb]
	\centering
	\includegraphics[width=1\textwidth]{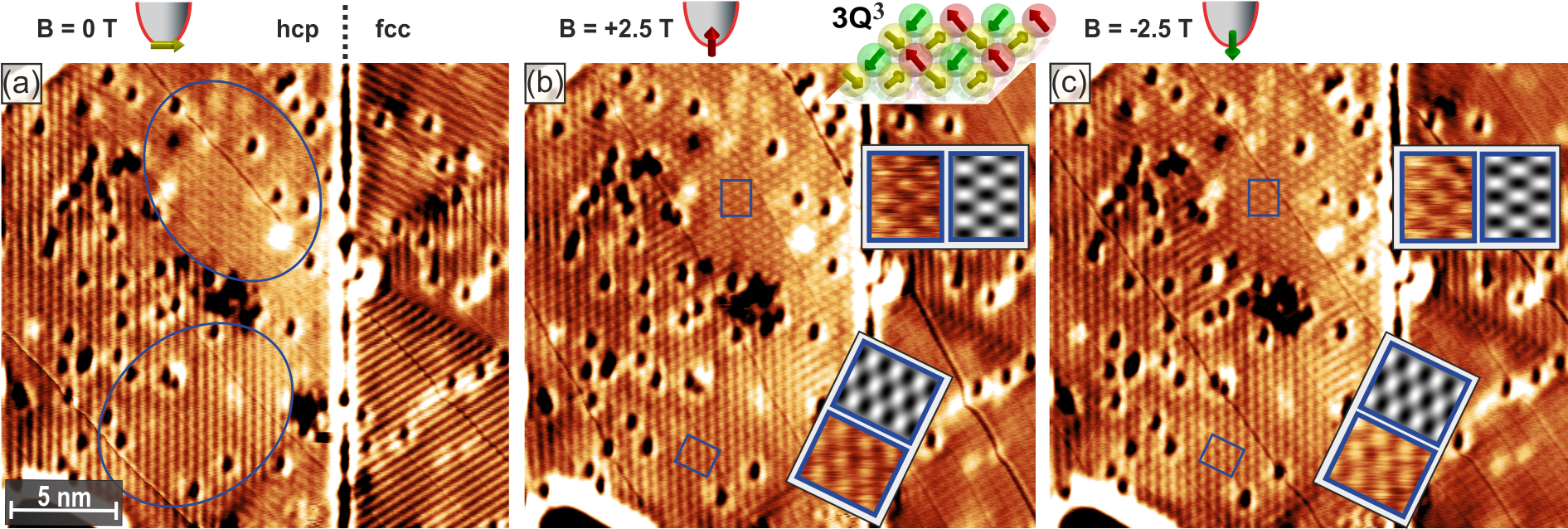}
	\caption{Spin-resolved STM images of hcp- and fcc-Mn areas with different tip magnetization directions sensitive to \textbf{(a)}~in-plane and \textbf{(b,c)} opposite out-of-plane sample magnetization components; Fe-coated W tip, $U=-30$\,mV, $I=7$\,nA, $T=8$\,K, partially differentiated constant-current images~\cite{suppMnRe}; straight diagonal lines are digital feedback artifacts. Insets in (b) and (c) show enlarged raw data views of the indicated areas together with SP-STM simulations (gray scale) of  the 3Q$^3$ state sketched above, where yellow spins are fully in-plane.}
	\label{fig4}
\end{figure*}

In the ideal 3Q state all adjacent spins span angles of $\tau=\arccos(-1/3)\approx 109.47^\circ$ (tetrahedron angle). 
In a monolayer, there can be different orientations of 3Q states with respect to the plane, and also different permutations of the atoms among the different sites.
Three highly symmetric versions, denoted 3Q$^i$, $i=1,2,3$, are depicted in the centers of Figs.\,3(b-d) where the color indicates the out-of-plane magnetization component.
They can be transformed into one another by rotating all spins, i.e.\ $\tau/2$ from 3Q$^1$ to 3Q$^2$ and 90$^\circ$ from 3Q$^2$ to 3Q$^3$.
To the sides are SP-STM simulations based on a simplified model~\cite{heinzeAPA2006} assuming opposite out-of-plane tip magnetization directions.
For 3Q$^1$ two rotational domains exist which cannot be distinguished with an out-of-plane tip, while an inverted pattern is observed when either tip or sample magnetization is inverted. Both orientations 3Q$^2$ and 3Q$^3$ are uniaxial and three rotational domains are expected. Upon a tip or sample magnetization inversion, cf.\ Figs.\,3(b-d), the magnetic pattern of these states is preserved but shows a phase-shift in the simulation.

Experimentally we can measure  different magnetization components at the same sample position exploiting the field dependence of an Fe-coated W tip, see schematics at the top of Fig.\,4. 
The spin moments of 3Q and RW-AFM state are fully compensated on the atomic scale and should therefore not react to moderate external magnetic fields.
Figure\,4(a) shows a sample area with hcp Mn monolayer on the left side and fcc Mn monolayer on the right side of a dislocation line, measured with an in-plane sensitive tip.
Two hcp Mn areas with qualitatively different patterns are indicated, which we interpret as different rotational domains of the 3Q state.
When the tip is sensitive to the out-of-plane magnetization component, see Fig.\,4(b,c), the observed magnetic pattern changes and both areas look more similar~\cite{suppMnRe}. Close inspection of the data reveals, that the two patterns exhibit a phase-shift upon tip magnetization inversion, and we find that the pattern shifts in different directions for the two magnetic domains. This is best seen in the insets in Fig.\,4(b) and (c), where the raw data from the indicated areas is compared to simulated SP-STM images (gray) of the 3Q$^3$ state depicted above.
While the 3Q$^1$ and 3Q$^2$ orientations are inconsistent with this data, cf.\ Fig.\,3(b,c), the agreement of raw data and simulation indicates that the 3Q$^3$ state is realized in hcp Mn.

The question arises which magnetic interaction couples the 3Q state to the lattice in this particular way. There is no energy difference between the different 3Q$^i$ states when considering Heisenberg exchange, the HOI, the MAE or the DMI. An estimation of the dipolar energy for these non-collinear configurations shows that it changes by only 0.01 meV/atom between the different 3Q$^i$ states. The value of the ASE in nearest-neighbor approximation can be obtained from the DFT calculations in Fig.\,\ref{fig2}(c) and for hcp Mn it is $J^\mathrm{ASE}_1=\Delta E / 4 = 0.2$\,meV, which is an order of magnitude stronger compared to fcc Mn. However, because the ASE maximizes for collinear spin configurations, it leads to an energy difference between the different 3Q$^i$ of only $0.07$\,meV per Mn atom. While the experimental observations point to the 3Q$^3$ state, 3Q$^1$ is slightly favored by both the dipolar interaction and the ASE. The estimated total energy difference between these states is about 0.08 meV/Mn atom, a factor of three smaller than for the two RW-AFM states considered for fcc Mn.

In the studied system of Mn/Re(0001) several competing magnetic interactions are sizable, giving rise to a situation where two states and their different orientations are nearly degenerate in the 
DFT calculations. In the experiments, however, for each stacking of the Mn layer we find a specific spin texture with a specific orientation and three symmetry equivalent rotational domains. The near degeneracy of states allows small effects like the stacking to determine the magnetic ground state in the experiment, but also indicates that not all relevant effects were taken into account by the present calculations. Firstly, rigid tetrahedron angles in the 3Q state might be an oversimplification: slight distortions potentially reduce the energy cost of MAE, ASE, and dipolar contributions and at the same time change the considered HOI energies, which might affect ground state energies and orientations.
Secondly, because the considered HOIs nearly cancel, additional higher order terms might play a decisive role in this system. Moreover, due to the large angles between adjacent atoms, the 3Q state carries a significant topological charge of $q=0.5$ per triangular plaquette or 2 per magnetic unit cell; for comparison, a skyrmion carries $q=1$. This can give rise to orbital moments~\cite{hankePRB2016} and energy contributions from chiral-chiral or spin-chiral interactions~\cite{grytsiukNC2020} with possible effects on magnetic ground states and their orientations.

We conclude by emphasizing that two model-type magnetic states have been found experimentally for the first time, i.e.\ the RW-AFM state and the 3Q state. They arise in the two different stackings of the Mn monolayer on Re(0001), and in both cases the orientation of the spin structure couples to the atomic lattice.
For the RW-AFM$_\|$ state the dipolar interaction and the anisotropic symmetric exchange can explain this coupling.
For the 3Q$^3$ state, we find that these two terms are too small to cause the coupling, and other effects such as a distortion of the spin state or contributions from orbital moments may be responsible. Complex spin structures such as the 3Q state are promising candidates to induce topological superconductivity \cite{nakosaiPRB2013}
below the critical temperature of about 1.7\,K of Re, and to exhibit interesting transport properties even in the normal conducting state.

\section{Acknowledgments} 
K.v.B.\ S.H., and A.K.\ acknowledge financial support from the Deutsche Forschungsgemeinschaft (DFG, German Research Foundation)-418425860;-402843438;-408119516, R.W.\ acknowledges financial support by the EU via the ERC Advanced Grant ADMIRE. K.v.B.\ thanks S.\,Bl{\"u}gel for pointing out valuable references. A.K.\ thanks L.\,Rosza for discussions and J.\,Sass\-mannshausen for technical support. S.H.\ and S.M.\ thank Y.~Mokrousov and G. Bihlmayer for valuable discussions, S.\,Haldar, S.\,Paul, and M.\,Goerzen for technical support and gratefully acknowledge computing time at the supercomputer of the North-German Supercomputing Alliance (HLRN).

\end{document}